\newcommand{\ls}{\underline{s}}
\newcommand{\De}{\Delta}

\newcommand{\vph}{\varphi}
\newcommand{\p}{\partial}

\documentclass[final]{aipproc}
\layoutstyle{6x9}
\usepackage{amsmath}
\usepackage{graphicx,bm}

\begin{document}

\title{Three-Dimensional PIC Simulation of Electron Plasmas}

\classification{}
\keywords{}

\author{M. Rom\'e}{
address={I.N.F.N., Dipartimento di Fisica, 
Universit\`a degli Studi di Milano, Italy}}

\author{R. Pozzoli}{
address={I.N.F.N., Dipartimento di Fisica,
Universit\`a degli Studi di Milano, Italy}}

\author{M. Pravettoni}{
address={I.N.F.N., Dipartimento di Fisica, 
Universit\`a degli Studi di Milano, Italy}}

\author{Yu. Tsidulko}{
address={Budker Institute of Nuclear Physics, Novosibirsk, 
Russian Federation}}

\begin{abstract}
The three-dimensional evolution of a pure electron plasma
is studied by means of a particle-in-cell code
which solves the drift-Poisson system where kinetic effects 
in the motion parallel to the magnetic field are taken into account. 
Different results relevant to the nonlinear dynamics of trapped 
plasmas and low-energy electron beams are presented.
\end{abstract}

%\date{\today}
\maketitle

\section{Introduction}
The evolution of an electron plasma in a Malmberg-Penning trap 
\cite{malmberg75} is studied by means of the particle-in-cell (PIC) 
code MEP \cite{jcp}.
The electron dynamics is analyzed in the frame of a guiding center
electrostatic approximation, where the velocity perpendicular to 
an externally applied uniform axial magnetic field is given by the 
electric drift, and kinetic effects in the motion parallel to the
magnetic field are taken into account.
The evolution of the system is followed within a conducting cylindrical 
surface of radius $R$ and length $L$ on which the (in general time-dependent) 
boundary conditions for the electrostatic potential are imposed.

The code is applied here to investigate two different situations 
(see Fig.~1). In the first case, the 
evolution of a traveling electron beam is considered. It is observed 
in particular how in a space-charge dominated regime a fraction of 
the electrons close to the axis is reflected back to the cathode 
while a high density annulus is formed inside the drift tube, which 
gives rise to the development of vortex-like structures.
In the second case, the injection phase of an electron plasma in a 
Malmberg-Penning trap is studied. Here, the electrons
enter in an initially empty trapping region, are reflected by a fixed 
potential barrier on the opposite side of the trap and come back to the 
cathode, thus interacting with new emitted electrons. 
It is shown how, in dependence of injected current and geometrical 
parameters, a virtual cathode may form close to the injection surface,
and a hollow electron column is formed in the trapping region. Longitudinal
kinetic effects are also investigated by varying the velocity distributions 
of the injected electrons. 
\begin{figure}[t]
\centering
\includegraphics[width=0.90\textwidth]{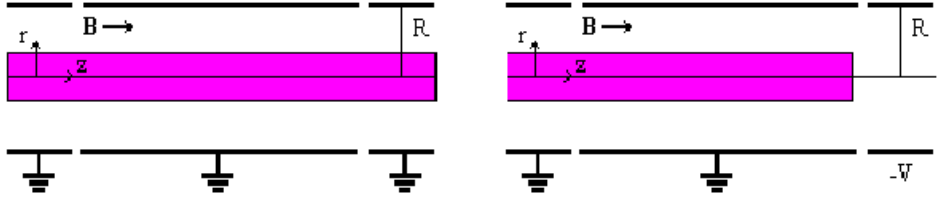}
\caption{Scheme of traveling beam configuration (\textbf{left})
and trap filling configuration (\textbf{right}). $-V$ is the
trap potential.}
\end{figure}
\section{Model and PIC code}
The system is described in the frame of the zeroth order
drift approximation, where the velocity perpendicular to the uniform 
magnetic field ${\bf B} = B {\bf e}_z$ (${\bf e}_z$ being the unit vector 
in the $z$ direction) is given by the electric drift, ${\bf{v}}_{E}=  
(c/B) \ {\bf e}_z\times \nabla \vph $, with $\vph$ the 
electrostatic potential and $c$ the speed of light. 
Assuming the guiding centers distribution, $f$, of the form $f({\bf{x}},
{\bf{v}},t)=F(v_{\|},{\bf{x}},t) \ \delta({\bf{v}}_{\bot} - {\bf{v}}_{E})$, 
where ${\bf{x}}, {\bf{v}},t$ are position, velocity and time, respectively, 
$\delta$ denotes the Dirac's distribution, and $v_\|$ and ${\bf{v}}_{\bot}$ 
denote the components of the velocity parallel and perpendicular to 
the magnetic field, the Vlasov-Poisson system reads
\begin{equation}
\frac{\p F}{\p t}+ (\frac{1}{2} \ {\bf e}_z \times \nabla \vph \
+v_\|{\bf e}_z ) \cdot \nabla F + 
\frac{1}{M_{eff}} \ {\bf e}_z \cdot \nabla \vph
\ \frac{\p F}{\p v_{\parallel}}  = 0 \quad ; \quad
\nabla^2\vph = n \, ,
\label{mep3e3}
\end{equation}
where $n({\bf x} , t) = \int F \, d v_{\parallel} $ is the electron 
density. Adimensional quantities are used:
length, time, density and potential are normalized over $R$,
$\omega_c/ 2 \omega_p^2$, $n_0$ and $4\pi e n_0 R^2$, respectively,
where $\omega_c \equiv e B / m c$ is the non-relativistic electron
cyclotron frequency, $\omega_p \equiv ( 4 \pi e^2 n_0 / m)^{1/2}$ 
is the electron plasma frequency, computed for a specified electron 
density $n_0$, and $- e$, $m$ are electron charge and mass, respectively.
The behavior of the system is therefore characterized by the single 
parameter $M_{eff}$, which plays the role of an effective mass, 
\begin{equation}
M_{eff} \equiv 4 \ \frac{\omega_p^2}{\omega_c^2} \ 
\simeq 4.115 \cdot 10^{-4}
\frac{n_0 [10^7 {\rm cm}^{-3}]}{B^2 [{\rm kGauss}]} \ .
\label{mep3e4}
\end{equation}
It results $M_{eff}= 2 n_0 / n_B$, $n_B$ being the so-called Brillouin 
density, $n_B \equiv (B^2/8 \pi)/m c^2$, so that $0 \leq M_{eff} \leq 2$.
Introducing a cylindrical system of coordinates $(r, \theta, z)$, and
using the variable $s \equiv r^2$, the equations of motion corresponding 
to the trajectories of the kinetic equation in (\ref{mep3e3}) are:
\begin{equation}
\frac{ds}{dt}= - \frac{\p\vph}{\p \theta} \quad ; 
\frac{d\theta}{dt}= \frac{\p\vph}{\p s} \quad ; 
\frac{d z}{dt}=v_{\parallel} \quad ; 
\frac{d v_{\parallel}}{dt}=\frac{1}{M_{eff}}\frac{\p\vph}{\p z} \, .
\label{mep3e5}
\end{equation}
In the code, Eqs.~(\ref{mep3e3}) are discretized on an equispaced grid 
(with the only exception of the central cell) 
in the coordinates $s$, $\theta$ and $z$. 
The number of cells is denoted as $N_s$, $N_\theta$ and $N_z$, 
respectively. The grid for $s$ is defined as 
$\ls_0=0,$ $\ls_1=1/(N_sN_\theta +1),$ 
$\ls_{j+1}=\ls_j+N_\theta/(N_sN_\theta +1),$ $s_0=0,$
$s_j=(\ls_{j}+\ls_{j+1})/2,$ $j=1\, ...\, N_s$ 
($s$ is the position of the center, while $\ls$ denotes the 
lower boundary of a ``radial'' cell). The grid for $\theta$ is 
$\theta_{l}=2\pi (l-1)/N_{\theta},$ $l=1\, ...\, N_{\theta},$ 
with the periodicity relation $\theta_{N_\theta +1}=\theta_1 $,
while the grid for $z$ is $z_k=(k-1/2)L/N_z- L/2,$ $ k=1\, ...\, N_z.$
Each cell has the same volume 
$\De V=\De s\De \theta \De z/2$, with $\De s=1/(N_s+1/N_\theta ),$ 
$\De \theta =2\pi/N_{\theta},$ and $\De z=L/N_z,$ respectively.
The system governed by Eqs.~(\ref{mep3e3}) is simulated numerically
as an {\it ensemble} of macro-particles with fixed sizes $\De s$, 
$\De \theta$ and $\De z$, using a PIC method \cite{jcp}. 

\section{Results}
The code is used to simulate the plasma dynamics in a Malmberg-Penning 
trap for different possible experimental settings.
The physical situation is determined by several geometrical and emission
parameters: the magnetic field strength $B$; the sizes $R$, $L$ and
the geometry of the emitting surface; the potentials which are imposed 
on cathode, anode and drift tube; the initial electron velocity 
distribution and the initial current distribution emitted by the source.
Malmberg-Penning traps usually use a spiral-wound tungsten 
filament for the injection \cite{amoretti03}. The MEP code is able to simulate
this initial spatial distribution of the electrons;
the effect of an accelerating grid is considered by suitably ``cutting'' 
the spiral along rows and columns of a given width. 
In addition, the code is able to take into account various initial 
velocity and current density distributions.
\begin{figure}[t]
\centering
\includegraphics[width=0.50\textwidth]{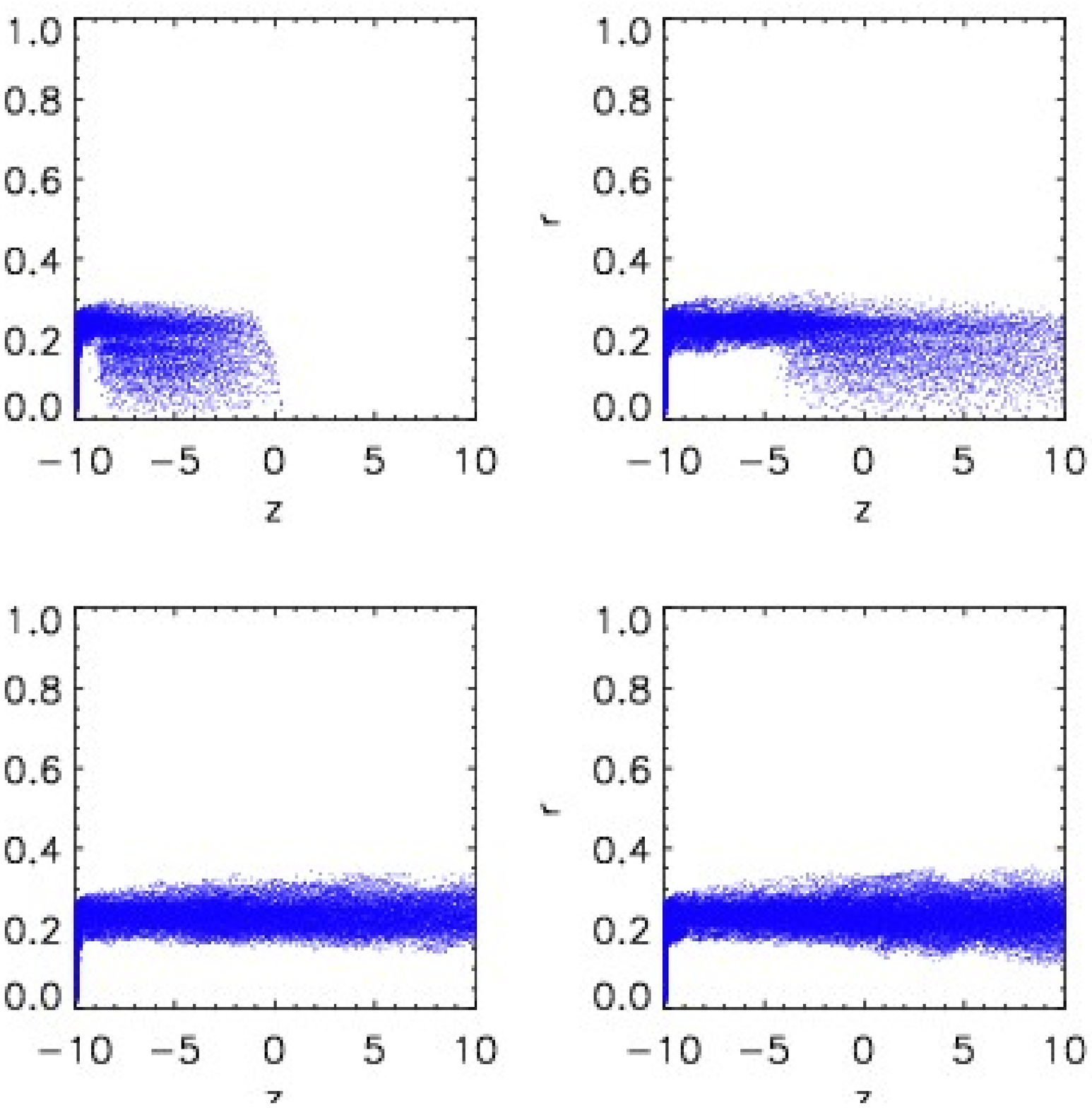}
\includegraphics[width=0.50\textwidth]{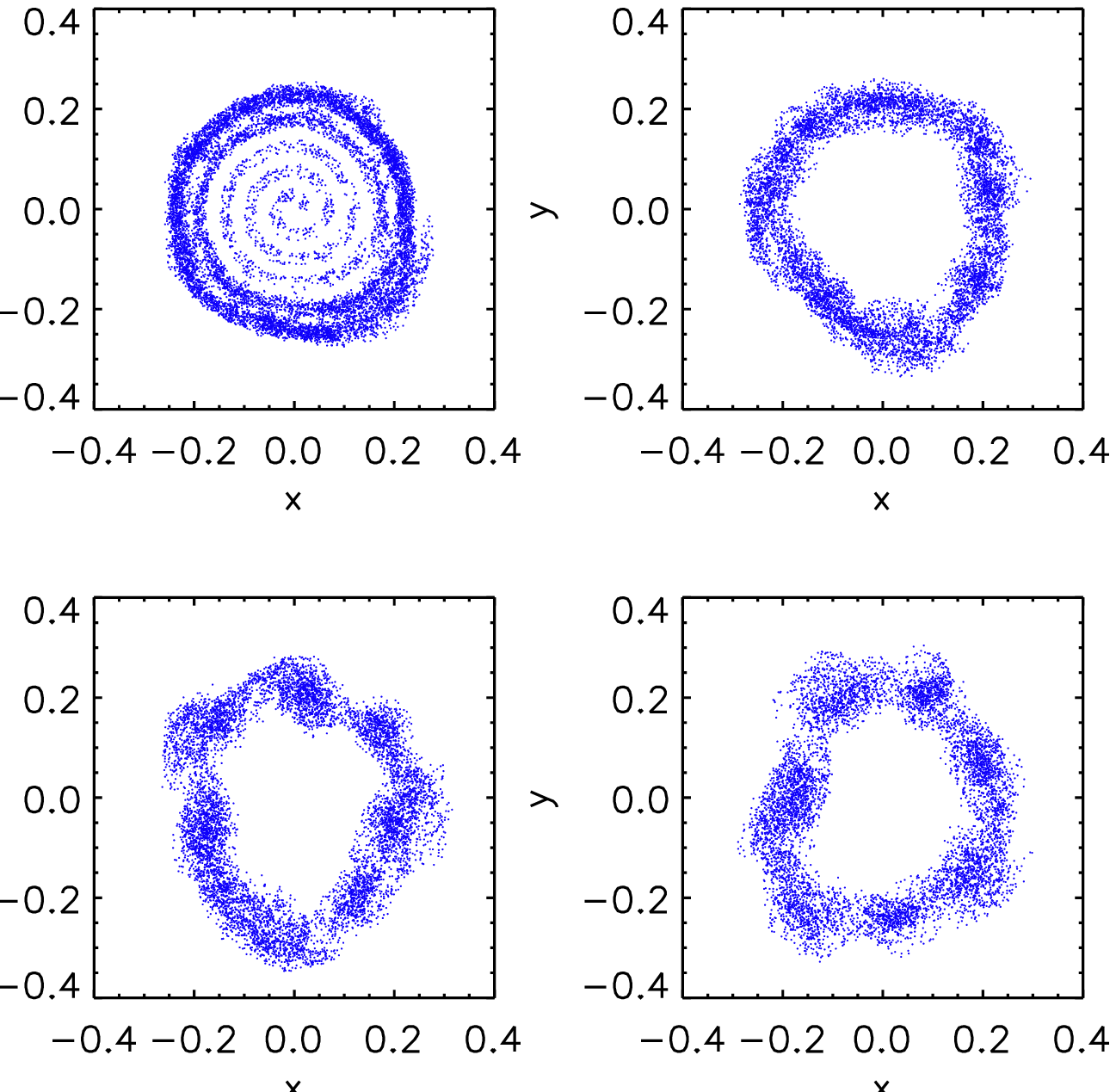}
\caption{{\bf Left:} Time evolution of a beam in the plane 
$(z,r)$. The parameters of the run are: $M_{eff}=0.01$,
$N_s=64$, $N_{\theta}=64$, $N_z=128$, $N_p=10^5$. The beam is 
injected at $z=-10$, with an initial parallel velocity $v_{\parallel 0} 
= 9.0$. The current is $I=0.2215$, $I$ being normalized over the 
ratio between $e \pi R^2 L n_0$ and the normalization time.
Emitter and collector are both at ground. From left to right, top to bottom, 
the data refer to $t = 0.7$, $1.5$, $3.5$ and $10.0$, respectively.
{\bf Right:} Transverse projection [$(x,y)$ plane]
of the particles in the interval $-10 \leq z \leq -5$,
$-5 \leq z \leq 0$, $0 \leq z \leq 5$ and $5 \leq z \leq 10$,
respectively. The data refer to $t=10.0$.}
\end{figure}
At first, a low-energy electron beam traveling in an equipotential
(grounded) drift tube is considered (see Fig.~1 {\it left}).
The electrons are continuously emitted from the cathode and collected 
to the anode. 
The characteristic time scale of the system is the time of flight of 
an electron, $L/v_{\|}$. In a space-charge-limited regime, 
it is found experimentally that the central part of the beam is 
reflected, a hollow electron column forms and fast coherent structures arise,
possibly due to the development of diocotron instability \cite{bettega04}. 
The PIC simulations confirm the experimental findings.
The time evolution of a mono-energetic beam in the $(r,z)$ plane 
is shown in Fig.~2 {\it left}. It is evident, in particular, the formation 
of a virtual cathode close to the injection surface. The central 
part of the beam is reflected back to the cathode by the 
space-charge of the beam itself, and only the outer part of the 
beam reaches the opposite end of the trap. This annular beam 
shows a quasi-2D evolution of vortex structures: 
Fig.~2 {\it right} represents the electron plasma distribution in 
(almost) stationary conditions on different transverse slices.

As a second example, the case of an electron plasma 
filling the trap is studied (see Fig.~1 {\it right}).
This situation simulates the phase of injection in a 
Malmberg-Penning trap. In this phase, the end plug electrode is 
maintained at a sufficiently negative potential in order to reflect 
the electrons entering the trap, while the plug electrode at the 
entrance is grounded to let the electrons flow into 
the trap from the cathode (starting from this configuration, the
trapping phase is obtained by simply lowering the plug potential
at the entrance of the beam to the same value as the end potential,
so that the electrons are electrostatically confined along the axis
of the device).
\begin{figure}[t]
\centering
\includegraphics[width=0.50\textwidth]{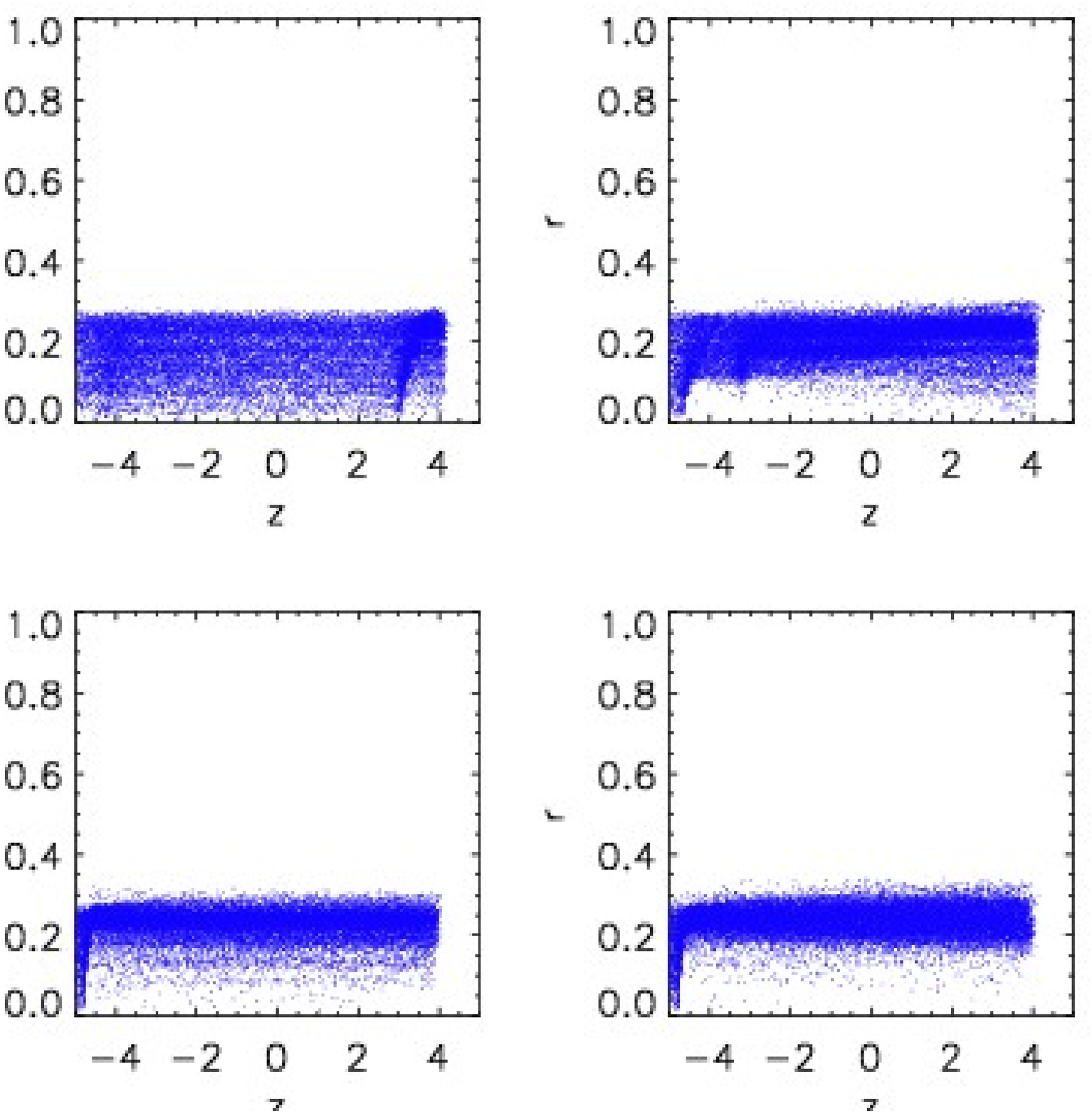}
\includegraphics[width=0.50\textwidth]{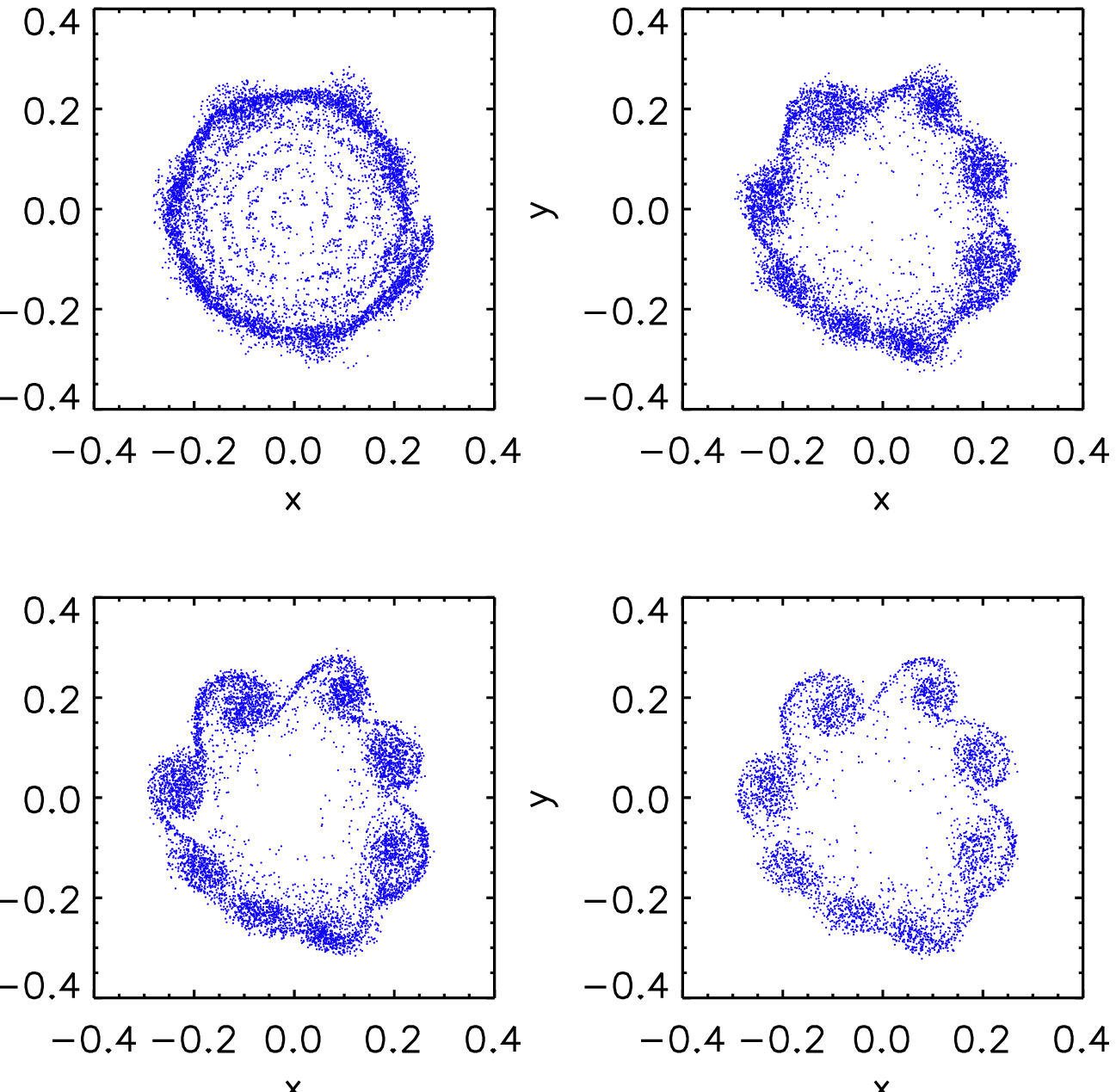}
\caption{{\bf Left:} Time evolution of a plasma filling the trap, 
in the $(z,r)$ plane. The parameters of the run are: $M_{eff}=4.64
\cdot 10^{-4}$,
$N_s=64$, $N_{\theta}=64$, $N_z=128$, $N_p=10^5$.
The beam is injected at $z=-5.0$, with $v_{\parallel 0} = 28.90$. 
The current is $I=0.1423$, and the potential difference
between collector and emitter is $-0.97$. From left to right,
top to bottom, the data refer to $t = 0.7$, $1.5$, $4.0$ and $8.0$, 
respectively. {\bf Right:} Transverse projection
of the particles in the interval $-5 \leq z \leq -2.5$,
$-2.5 \leq z \leq 0$, $0 \leq z \leq 2.5$ and $2.5 \leq z \leq 5$,
respectively. The data refer to $t=8.0$.}
\end{figure}
A mono-energetic beam is considered. It is found that when the input 
current of the beam is increased, the space charge cloud which forms 
inside the trap represents a barrier not only for the new incoming 
particles, but also for the electrons which are already inside
and are reflected by the external negative potential barrier.
As a result, the electron plasma filling the trap assumes
an annular shape, as shown in Fig.~3. The time evolution
of the electrons in the phase plane $(z,p_{\parallel})$ is shown
in Fig.~4 (red dots), where $p_{\parallel} \equiv M_{eff} 
v_{\parallel}$. Longitudinal kinetic effects have been 
investigated in this case. In particular, it has been found that  
varying the parallel velocity distribution of the injected 
electrons to a Maxwellian has a dramatic effect (see again Fig.~4, 
blue dots): space-charge effects are much weaker in this case, 
and the electron plasma maintains its initial transverse (spiral) shape.
\begin{figure}
\centering
\includegraphics[width=0.90\textwidth]{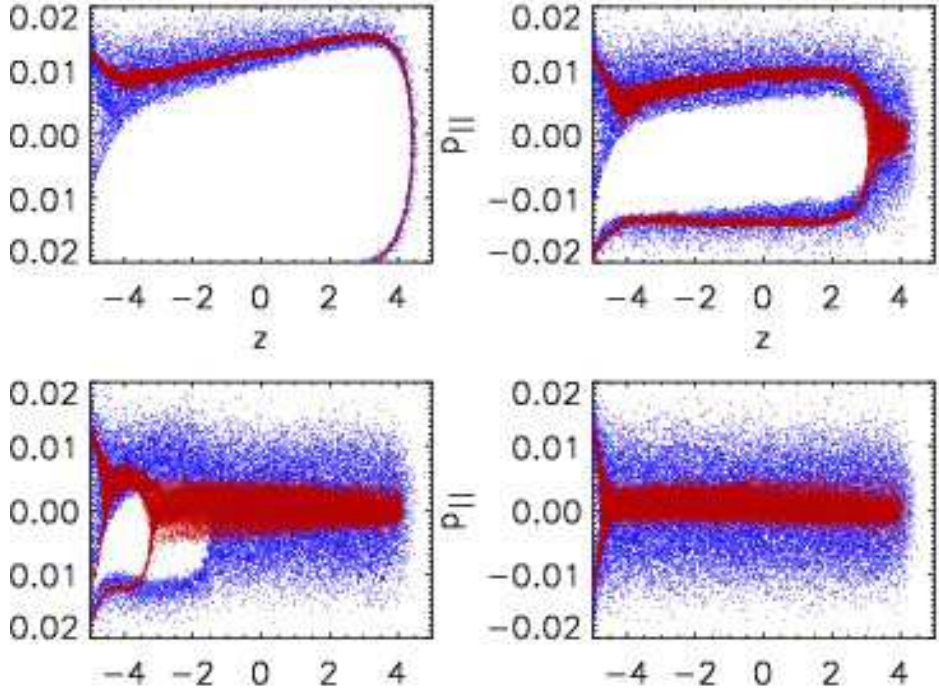}
\caption{Time evolution of a plasma filling the trap,
in the $(z,p_{\parallel})$ plane.  
The parameters are the same as in Fig.~3 (red dots). 
From left to right,
top to bottom, the data refer to $t = 0.3$, $0.7$, $1.5$ and $3.0$,
respectively. The blue dots are relevant to an initial Maxwellian 
distribution in parallel velocity, $F(v_{\|},{\bf{x}},0) = 
g({\bf x}) \cdot \exp [-(v_{\|}-v_{\|0})^2/2\sigma^2),$
with the same central value $v_{\|0}$, and a velocity spread 
$\sigma = 5.78$.}
\end{figure}


\begin{thebibliography}{99}
\bibitem{malmberg75}
J. H. Malmberg and J. S. de Grassie,
Phys. Rev. Lett. \textbf{35}, 577 (1975).

\bibitem{jcp}
Yu. Tsidulko, R. Pozzoli and M. Rom\'e,
submitted to J. Comp. Phys. (2004).

\bibitem{amoretti03}
M. Amoretti, G. Bettega, F. Cavaliere, M. Cavenago, 
F. De Luca, R. Pozzoli and M. Rom\'e, 
Rev. Scient. Instr. \textbf{74}, 3991 (2003).

\bibitem{bettega04}
G. Bettega, F. Cavaliere, M. Cavenago, A. Illiberi,
R. Pozzoli, M. Rom\'e and Yu. Tsidulko,
Appl. Phys. Lett. \textbf{84}, 3807 (2004).
\end{thebibliography}
\end{document}